\documentclass[prl,twocolumn,floats]{revtex4}
\usepackage{graphicx,epsfig}

\begin{document}


\title{Quantization of the diagonal resistance: Density gradients and the empirical resistance rule in a 2D system}

\author{W. Pan$^{1,2,3}$, J.S. Xia$^{3,4}$, H.L. Stormer$^{5,6}$, D.C. Tsui$^2$, C.L. Vicente$^{3,4}$,\\
E.D. Adams$^{3,4}$, N.S. Sullivan$^{3,4}$, L.N. Pfeiffer$^6$, K.W. Baldwin$^6$, and K.W. West$^6$}

\affiliation{$^1$Sandia National Laboratories, Albuquerque, New Mexico 87185}
\affiliation{$^2$Princeton University, New
Jersey 08544}
\affiliation{$^3$National High Magnetic Field Laboratory, Tallahassee, Florida 32310}
\affiliation{$^4$University of Florida, Gainesville, Florida 32611}
\affiliation{$^5$Columbia University, New York, New York
10027}
\affiliation{$^6$Bell Labs, Lucent Technologies, Murray Hill, New Jersey 07954}

\vskip5pc

\begin{abstract}

We have observed quantization of the {\it diagonal} resistance, $R_{xx}$, at the edges of several quantum Hall states. Each
quantized $R_{xx}$ value is close to the difference between the two adjacent Hall plateaus in the {\it off-diagonal}
resistance, $R_{xy}$. Peaks in $R_{xx}$ occur at different positions in positive and negative magnetic fields. Practically
all $R_{xx}$ features can be explained {\it quantitatively} by a ~1\%/cm electron density gradient. Therefore, $R_{xx}$ is
determined by $R_{xy}$ and unrelated to the diagonal resistivity $\rho_{xx}$. Our findings throw an unexpected light on the
empirical resistivity rule for 2D systems.

\end{abstract}


\date{\today}
\maketitle

The quantum Hall effects \cite{qhebook1,qhebook2,qhebook3} in a two-dimensional electron system (2DES) in a high magnetic
field ($B$) are characterized by plateaus quantized to integer or fractional values of the resistance quantum, h/e$^2$, in
the off-diagonal resistance, $R_{xy}$, with concomitantly occurring vanishing values in the diagonal resistance $R_{xx}$. A
multitude of integer (IQHE) and fractional (FHQE) quantum Hall effects have been discovered over the past two decades,
pointing to a wide variety of underlying single-particle and many-particle electronic states. In all cases, resistance
quantization is the purview of $R_{xy}$, while vanishing resistance occurs solely in $R_{xx}$.

In general the Hall resistance, $R_{xy}$, is a monotonic function of $B$ following a somewhat irregular sequence of slightly
broadened steps. The magnetoresistance, $R_{xx}$, has the appearance of the derivative of $R_{xy}$ with respect to $B$. In
fact, a phenomenological relationship, $R_{xx} \propto B \times dR_{xy}/dB$, called the resistivity rule, is so closely
observed in experiment \cite{chang85,stormer,rotger} that one suspects a deep fundamental relationship. While there exists a
somewhat complex model based on density fluctuations that can account for the general features \cite{simon} the origin of the
empirical resistivity rule continues to be enigmatic.

Recent experiments on ultra-high quality specimens performed at very low temperatures in the second Landau level
\cite{eisenstein,xia} have yielded data that seem to contradict the resistivity rule. Instead of rising in a stair-like
fashion from the IQHE through the FQHE plateaus to the next IQHE, Rxy switches back and forth several times between FQHE and
IQHE values. This phenomenon is called the reentrant IQHE or RIQHE. At present we have only an ad hoc understanding of its
origin \cite{shibata,goerbig}. As a result of the RIQHE, $R_{xy}$ is no longer monotonic and the empirical resistivity rule
now implies regions of negative $R_{xx}$, which is contrary to simple inspection. It is unclear, whether another relationship
holds in this region.

In examining the relationship between $R_{xx}$ and $R_{xy}$ in this filling factor regime, we discovered at temperatures of
~6mK, regions of {\it quantized diagonal-resistance}, $R_{xx}$. The quantized values of $R_{xx}$ are very close to the
differences of the two adjacent Hall plateaus in $R_{xy}$ and we can explain them in terms of a slight, unintentional
electron density gradient. Furthermore, we can explain {\it quantitatively} the complete $R_{xx}$ trace, including its
asymmetry in $B$ field directions, as being the result of such a density gradient. In turn, the resistivity rule derives
trivially from such a gradient picture. In this way, $R_{xx}$ is just a reflection of $R_{xy}$ measured at two, slightly
different densities and the diagonal resistivity, $\rho_{xx}$, has practically no bearing on $R_{xx}$. We discuss the
implication of these astonishing findings.

Our experiments were carried out in a demagnetization/dilution refrigerator combination described in Ref. [9]. The lowest
temperature ($T$) in our experiment was $\sim$ 6~mK. The 2DES is confined in a 30nm wide GaAs/AlGaAs quantum well that is
delta-doped on both sides of the well at a setback distance of 100~nm. The wafer was grown under rotation to minimize
electron density gradients. The specimen consists of a 4~mm $\times$ 4~mm square with 8 diffused indium contacts as seen in
the inset of Figure 2. The electron density is $n=3\times10^{11}$ cm$^{-2}$ and the mobility is $\mu=31\times10^6$ cm$^2$/Vs.
Lock-in techniques with $\sim$ 7~Hz were utilized to measure $R_{xx}$ and $R_{xy}$ at an excitation current of 1~nA.

Figure 1 shows our data on $R_{xx}$ and $R_{xy}$ at $T \sim$ 9mK. The top trace in light gray shows the Hall resistance,
$R_{xy}$, in the negative $B$ field direction ($B-$). $R_{xy}$ in the positive $B$ field direction ($B+$) is virtually
identical to this $R_{xy}$ and not shown. The $R_{xy}$ trace shows clear FQHE plateaus at $\nu=5/2, 7/3, 8/3$. Four RIQHE
states are apparent. At these positions $R_{xy}$, for a finite $B$-field region, assumes an IQHE value. The bottom part of
Fig. 1 shows $R_{xx}$ data. $R_{xx}(B+)$ is plotted in black showing the expected deep minima at the positions of plateaus in
$R_{xy}$ and sharp spikes at their edges. However, in this case the spikes are concentrated on the sides at which $R_{xy}$ is
{\it rising} sharply, whereas the sides associated with sharp {\it drops} in $R_{xy}$ are almost absent in $R_{xx}$. Fig. 1
also shows $R_{xx}(B-)$ data but plotted as $-R_{xx}(B-)$, in the negative $y$-direction, for later comparisons. As in
$R_{xx}(B+)$ we observe the required minima and the sharp spikes at their flanks. In this case, however, the spikes are
concentrated on the sides at which $R_{xy}$ is {\it dropping} sharply.

To examine the resistivity rule in the second Landau level, we show in Fig. 1 $R_{diff}=B\times dR_{xy}/dB$, as a gray trace,
multiplied by a constant $c=0.003$ for comparison with $R_{xx}$. There is a very strong resemblance between the traces. The
positive parts of $c \times R_{diff}$ are well reflected in $R_{xx}(B+)$, whereas the negative going parts of $R_{diff}$ do
not find an equivalent reflection in $R_{xx}$. On the other hand, the negative going sections of $c\times R_{diff}$ closely
match the inverted $R_{xx}(B-)$ trace. At first such an asymmetry appears puzzling but will make sense in the context of the
following $R_{xx}$ quantization and the edge channel picture.

In Fig.2, we show $R_{xx}(B+)$ as a black trace taken at $T \sim$ 6~mK. The main features are similar to what was observed at
$T \sim$ 9~mK in Fig. 1. However, at 6mK, when the Hall resistance varies abruptly, $R_{xx}$ assumes a flat top. This is
clearly seen in the upper inset to Fig. 2, which shows a magnification of the region Q1, near $\nu=8/3$. A similar
quantization is observed in the region marked Q2 at a $B$ field $\sim$ 5.58~T and close to $\nu=7/3$. $R_{xy}$ (thin light
gray trace), on the other hand, shows no anomaly but varies monotonically between plateaus. The observed $R_{xx}$ plateaus
appear only for $T < 9$~mK and their width increase with decreasing $T$.

The $R_{xx}$ plateau values are very close to the difference between the values of the two adjacent quantum Hall plateaus in
$R_{xy}$. For example, around $\nu=8/3$, $R_{xx} \cong$ 1097~ohms, while $\Delta R_{xy} = R_{xy}(\nu=8/3)-R_{xy}(\nu=3) =
h/e^2 \times (3/8-1/3) \sim$ 1076~ohms. The solid dots in Fig. 2 represent the value of $\Delta R_{xy}$ calculated from the
adjacent $R_{xy}$ pairs. It is apparent that $R_{xx} \cong \Delta R_{xy}$ holds for all well developed peaks, even for those
between adjacent FQHE states, such as between $\nu=2+2/5$ and 2+3/8 and between $\nu=2+3/8$ and 2+1/3.

Having made this discovery in the second Landau level, we tested the relationship more generally and found that $R_{xx} \cong
\Delta R_{xy}$ also holds for the well-developed peaks in the lowest Landau level. Fig. 3 shows $R_{xx}$ (black) and $R_{xy}$
(light gray) traces for $2 > \nu > 1$. The solid dots represent $\Delta R_{xy}$. Their position, in many cases, matches
closely the height of the strong peaks in $R_{xx}$, although, in this filling factor range we did not observe quantization of
$R_{xx}$.

Plateaus in $R_{xx}$ have been previously observed in macroscopically inhomogeneous 2D systems
\cite{syphers,buttiker,washburn,komiyama,dorozhkin,chang92,merz,haug}. In fact, the phenomenology of these data is
reminiscent of the content of Figs. 1 and 2. For example, in Si-MOSFETs with spatially varying electron densities Syphers and
Stiles \cite{syphers} observed $R_{xx}$ quantization and strong peaks in $R_{xx}$ in one $B$-field direction, but practically
vanishing $R_{xx}$ in the opposite field direction. In all cases, the observations are best explained in the edge channel
picture. At the boundary between two regions of different filling factors $\nu_1$ and $\nu_2$ ($\nu_1 > \nu_2$), a total of
$\nu_1-\nu_2$ edge channels are being reflected. This leads to a voltage drop between these two regions equivalent to $R_{xx}
= h/e^2 \times (1/\nu_2-1/\nu_1)$. From the similarities between our data and those reported in the literature, we deduce
that a very small, unintentional density gradient exists in our 2D specimens. The width $\Delta B \sim 0.01$~T of the Q1
translates directly into a difference of $\sim$ 0.5\% in density between the requisite voltage probes ($V_{xx}: 4 \to 7$,
lower inset Fig. 2). Having a distance of about 5~mm, this represents an ~1\%/cm density gradient.

Since the $R_{xx}$ plateaus seem to be caused by a density gradient one wonders whether other $R_{xx}$ features of Fig. 2
follow. To test this, we assume a density $n_1$ at voltage probe 4 and a density $n_2$ at voltage probe 7 and calculate the
difference $\Delta R_{xy} =R_{xy}(n_1)-R_{xy}(n_2)\equiv R_{xy}(n)-R_{xy}(n+\Delta n)$ with $\Delta n=n_2-n_1$. This is
performed numerically, by subtracting the 6~mK $R_{xy}$ trace from the same trace with a $B$-field axis compressed by
$2\Delta n/(n_1+n_2) \cong 0.5$\%. The resulting data are plotted as a thick gray trace in Fig. 2. The resemblance with
$R_{xx}$ is striking. The positive part of $\Delta R_{xy}$ reproduces $R_{xx}(B+)$ perfectly, almost over the entire region.
The negative part of $\Delta R_{xy}$, on the other hand, closely matches the inverted $R_{xx}(B-)$. Most importantly, this
procedure creates the quantized $R_{xx}$ plateaus of Fig. 2. In fact, as in the second Landau level, $\Delta R_{xy}$ (dotted
line in Fig.~3) for a density difference of 0.5\% also reproduces $R_{xx}$ between $\nu=1$ and $\nu=2$ {\it quantitatively}.
An exception arises around $B \sim 11$~T, where the $\Delta R_{xy}$ peak exceeds $R_{xx}$ by about a factor of 2.

It is surprising that a simple modelling based on density inhomogeneity can produce $R_{xx}$ almost perfectly from $R_{xy}$.
In fact, a density gradient can explain the origin of the entire resistivity rule $R_{xx}=c\times R_{diff}$ simply in terms
of $\Delta R_{xy}$. A small density difference $\Delta n$ between $R_{xx}$ voltage probes creates at any given $B$ a
difference in Hall voltage $\Delta V_{xy}(n,B)=I\times(R_{xy}(n,B)-R_{xy}(n+\Delta n,B)) \cong I \times dR_{xy}(n,B)/dn
\times \Delta n = I \times dR_{xy}(n,B)/dB \times B/n \times \Delta n = I \times B \times dR_{xy}(n,B)/dB \times (\Delta
n/n)$. The latter equality holds since $dB/B=dn/n$. Therefore, $\Delta R_{xy}=\Delta V_{xy}/I= B \times dR_{xy}(n,B)/dB
\times(\Delta n/n)= c \times R_{diff}$. Since $R_{xx} =c \times R_{diff}$ is borne out in experiments, the resistivity rule
can simply be derived from a density gradient. Most importantly, the adjustable parameter, $c$, is now determined as $c=
\Delta n/n$. The value $c=0.003$ we used in Fig.1 is in fair agreement with the density gradient of 0.005, considering that
the two sets of data were taken in two different cool-downs and with different illumination history.

Before discussing the implication of our findings we address the $B$-field polarity dependence seen in Figs. 1 and 2. This is
best accomplished in the edge channel picture. The inset to Fig.3 shows two idealized sample configurations (S1 and S2) with
current probes $I_1$, $I_2$ and voltage probes $V_1$ through $V_4$. Each sample contains two different quantum Hall regimes
caused by density inhomogeneity. For simplicity, we adopt QHE states $i=1$ and $i=2$. We first emphasize that the edge
channel structure does not depend on current direction, but only on the direction of the $B$-field, which is pointing into
the plane. The whole specimen is encircled by the $i=1$ edge state, whereas only a part of it contains a second $i=2$ edge
channel. The edge channels represent equipotentials. When an external current is applied, so-called ``hot spot''
\cite{girvin} develop (shown as black squares), at which the potential along the edge suddenly jumps. The $V_1 \to V_2$
voltage drop in S1 reflects a quantized value of $R_{xx}$ although for a different set of $\nu$'s than seen in Fig. 2. The
opposite side of the specimen ($V_3, V_4$) is in the zero-resistance state \cite{fischer}. If we reverse the ordering of
$i=1$ and $i=2$, as in S2, the hot spot switches sides and occurs between $V_3$ and $V_4$ while $V_1-V_2$ vanishes.

When the $B$-field is swept through a normal stretch of monotonically rising $R_{xy}$ values, the region of higher quantum
number always remains on the same side of the region with lower quantum number. Therefore, the hot spot will remain on the
same side of the sample and the density gradient is observed in $R_{xx}$ on that side only. As the $B$ field is swept through
anomalous regions of $R_{xy}$, in which $R_{xy}$ suddenly {\it drops} (as in the vicinity of RIQHE states) the spatial order
of quantum numbers is {\it reversed}, (as shown in S2) the hot spot switches sides, the voltage drop appears in $V_3-V_4$,
and $V_1-V_2$ enters the zero resistance state. Therefore, there are no $R_{xx}$ features observed between probes $V_1$ and
$V_2$. They rather appear on the opposite side of the sample. With a limited number of good contacts to our sample we could
not test this switching to the opposite side. However, $B$-field reversal has the same effect as reversal of the spatial
sequence of quantum numbers. Therefore, under reversed $B$-field, we observe at the $V_1, V_2$ contacts only features in
$R_{xx}$ when $R_{xy}$ experiences an anomalous, sudden {\it drop}. For normal sudden {\it rises} in $R_{xy}$ and this
$B$-field direction the hot spot resides on the opposite side of the sample and, hence, $V_1-V_2$ shows the zero resistance
state and no spike.

The edge channel model, combined with the existence of a density gradient of $\sim$ 1\%/cm, explains all our observations in
$R_{xx}$: quantization of $R_{xx}$, $R_{xx}$ as a reflection of $\Delta R_{xy}$, as well as the appearance of selected peaks
in $R_{xx}$ depending on $B$-field direction.  We emphasize that almost all features of $R_{xx}$ are explained {\it
quantitatively} in terms of $\Delta R_{xy}$ caused by a slight density gradient. This explains trivially the resistivity rule
in our specimen, not only in the second Landau level with its particularities, but also in the lowest Landau level.
Therefore, in our experiments, $R_{xx}$ does not provide any information on $\rho_{xx}$, but is only a reflection of
$R_{xy}$.

The irrelevance of $\rho_{xx}$ to $R_{xx}$ is rather disturbing. $R_{xx}$ has been used to study many QHE properties from the
size of the energy gap, to two-parameter scaling, to the observation of critical points and quantum phase transitions
\cite{qhebook1,qhebook3,wei88,jiang93}. To the degree to which interpretations of the data in terms of $R_{xx}$, $\Delta
R_{xy}$ or $c\times B \times dR_{xy}/dB$ lead to identical physics, our results do not impact such interpretations based on
$R_{xx}$. While we are unable to review all previous usages of $R_{xx}$, it appears unlikely that $\rho_{xx}$ is irrelevant
to all of them. Furthermore, the interpretation of our data rests on the edge channel picture, which is expected to hold well
only in the non-dissipative regime. It is, hence, surprising that even the data of Fig.3 can be interpreted in terms of this
model. While it has been shown that the transport at the Fermi level involves all electronic level below E$_f$ \cite
{pruisken}, it is not obvious that such a relationship holds globally in the presence of macroscopic inhomogeneities, such as
a gradient.

All our data were taken below 20~mK and the gradient model holds for the entire temperature range. In a different, high
mobility quantum well sample, this model holds at a high temperature of 1.2K \cite{unpublished}. In fact, the resistivity
rule has been pursued to temperatures as high as 100~K \cite{rotger} and one wonders whether the same density gradient is
responsible or whether another phenomenon enters. If the latter were the case, one wonders how the resistivity rule can hold
without modification or transition for this entire temperature range; starting at low temperatures with a density gradient as
its origin, to the highest temperature, with a totally different origin, probably related to $\rho_{xx}$.

We thank E. Rezayi, C. de Morais Smith, M.O. Goerbig, S.M. Girvin, V. Scarola, S. Simon, and C.G. Zhou for discussions. This
work was supported by NSF under No. DMR-03-52738 and DOE under No. DE-AIO2-04ER46133. The work at Princeton was supported by
the AFOSR, the DOE, and the NSF. The work at Columbia was supported by DOE, NSF, and by the W.M. Keck Foundation. Sandia is a
multiprogram laboratory operated by Sandia Corporation, a Lockheed-Martin company, for the U.S. Department of Energy under
Contract No. DE-AC04-94AL85000. Experiments were carried out at the high $B/T$ facilities of the NHMFL, which is supported by
NSF Cooperative Agreement No.DMR-0084173, the State of Florida, and the DOE.

\begin{figure} [h]
\centerline{\epsfig{file=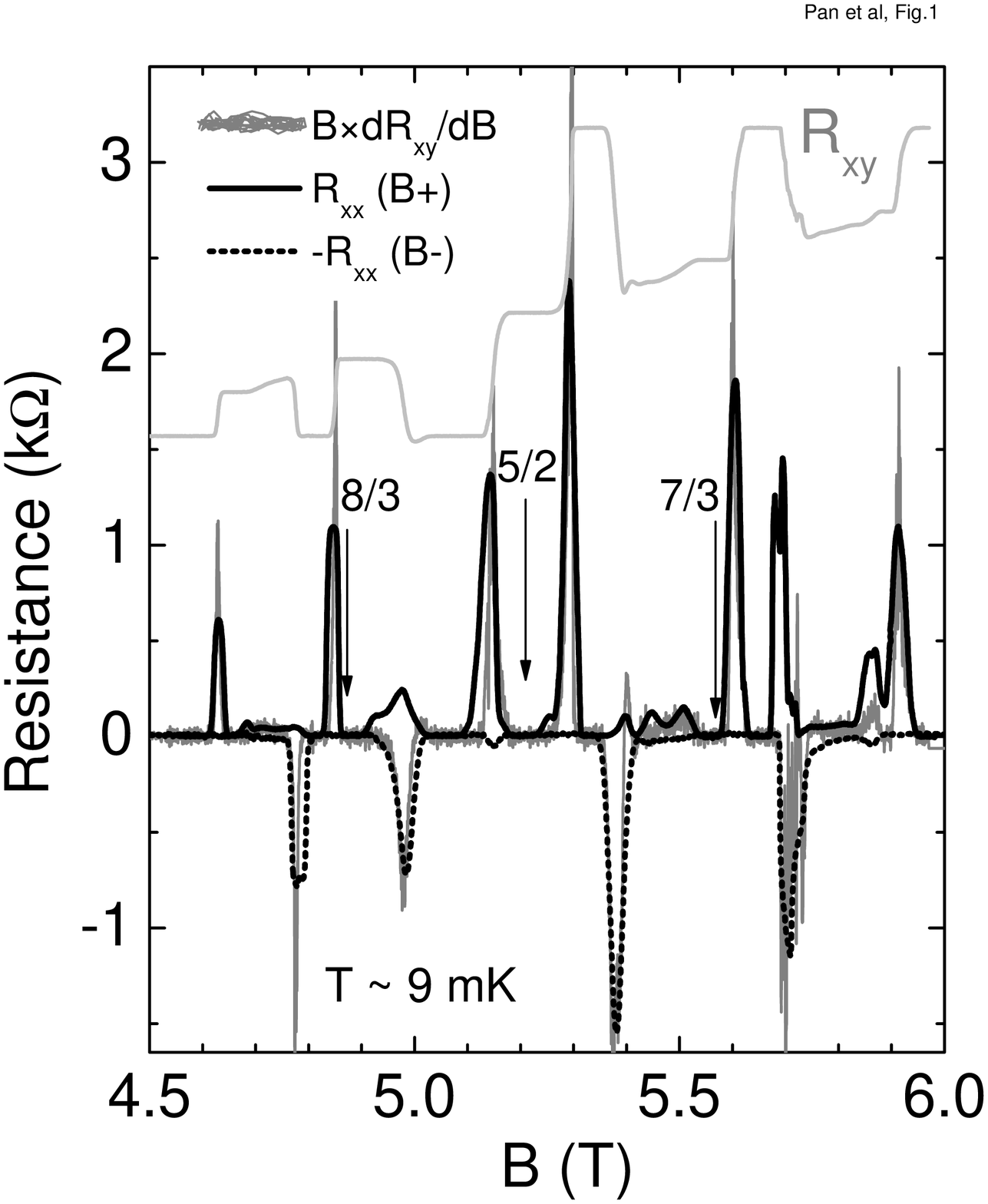,width=12.5cm}} \caption{Comparison of $B \times dR_{xy}/dB$ (noisy gray trace), and
$R_{xx}(B+)$ (black), $-R_{xx}(B-)$ (dotted line). The Hall resistance in the reversed $B$ field direction is plotted as the
light gray trace. Arrows mark the positions of prominent FQHE states.}
\end{figure}

\begin{figure} [h]
\centerline{\epsfig{file=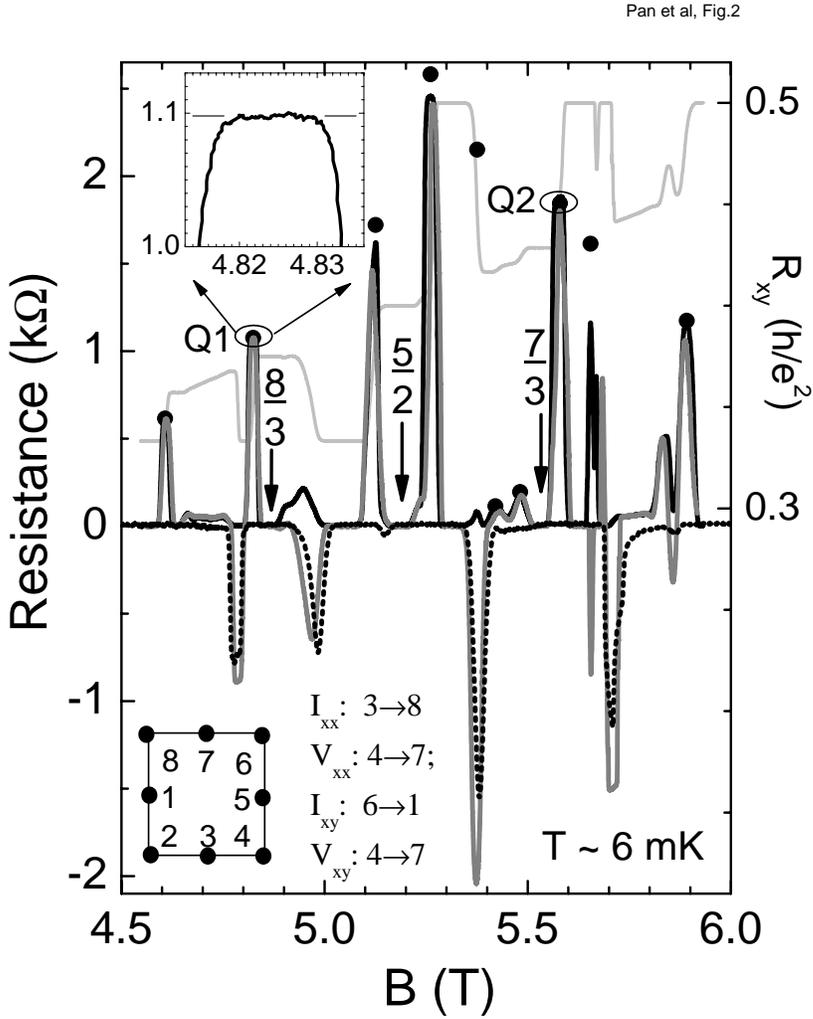,width=12.5cm}} \caption{Main plot: $R_{xx}(B+)$ (black trace), $-R_{xx}(B-)$ (dotted line),
and $\Delta R_{xy} = R_{xy}(n) - R_{xy}(n-0.005n)$ (gray curve), where $n$ is electron density. The temperature is $\sim$
6~mK. Arrows mark the major FQHE states. The bottom inset shows contact configuration for $R_{xx}$ and $R_{xy}$ measurement.
In the top inset, the region marked as Q1 is magnified to show the quantization of $R_{xx}$.}
\end{figure}

\begin{figure} [h]
\centerline{\epsfig{file=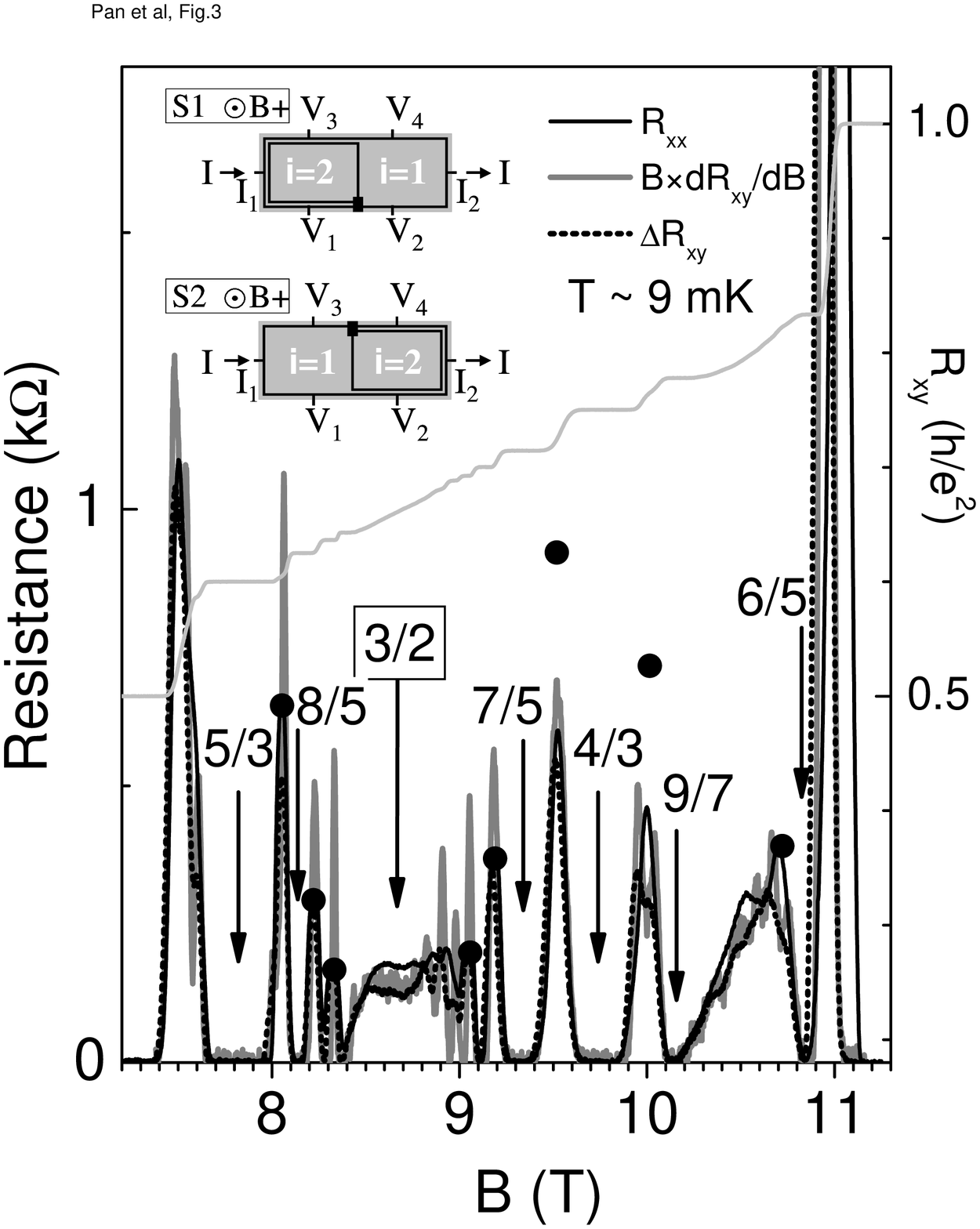,width=12.5cm}} \caption{Magnetoresistance in the first landau level around $\nu=3/2:
R_{xx}(B+)$ (black trace), $B \times dR_{xy}/dB$ (gray curve), and $\Delta R_{xy} = R_{xy}(n)- R_{xy}(n+0.005n)$ (dotted
line). The inset shows the edge model with density inhomogeneity. See text for details.}
\end{figure}

\end{document}